\documentclass{article}
\usepackage{iclr2027_conference,times}

\usepackage{amsmath,amsfonts,bm}

\def\eqref#1{equation~\ref{#1}}

\def\1{\bm{1}}

\DeclareMathAlphabet{\mathsfit}{\encodingdefault}{\sfdefault}{m}{sl}
\SetMathAlphabet{\mathsfit}{bold}{\encodingdefault}{\sfdefault}{bx}{n}

\usepackage{hyperref}
\usepackage{url}
\hypersetup{
  hidelinks,
  pdftitle={EasyFoam: Skill-Oriented OpenFOAM Agents with Evidence-Driven Evolution},
  pdfauthor={Peng Yin, Zhiwei Chen, Tianyi Liu, Chen Li, Kai Li, Jian Cheng}
}
\usepackage{booktabs}
\usepackage{multirow}
\usepackage{amsmath,amssymb}
\usepackage{graphicx}
\usepackage{xcolor}
\usepackage{tikz}
\usetikzlibrary{arrows.meta,positioning,fit,shapes.geometric,calc}
\usepackage{tabularx}
\usepackage{array}
\newcolumntype{Y}{>{\raggedright\arraybackslash}X}

\title{EasyFoam: Skill-Oriented OpenFOAM Agents with Evidence-Driven Evolution}
\author{
\textbf{Peng Yin}$^{1}$\thanks{Equal contribution.}
\quad
\textbf{Zhiwei Chen}$^{2,1}$\footnotemark[1]
\quad
\textbf{Tianyi Liu}$^{3}$
\\
\textbf{Chen Li}$^{3}$
\quad
\textbf{Kai Li}$^{1}$\thanks{Corresponding author.
Email: \texttt{kai.li@ia.ac.cn}}
\quad
\textbf{Jian Cheng}$^{1}$
\\
$^{1}$Institute of Automation, Chinese Academy of Sciences
\\
$^{2}$Nanjing University
\\
$^{3}$State Key Laboratory of Aerodynamics
\\
}

\iclrfinalcopy

\begin{document}

\maketitle
\fancyhead{}
\renewcommand{\headrulewidth}{0pt}

\begin{abstract}
Large language model (LLM) agents are increasingly capable of automating end-to-end OpenFOAM workflows for computational fluid dynamics (CFD). However, current progress is bottlenecked by two critical limitations. First, existing evaluations are dominated by tutorial-level cases and execution-oriented metrics, masking a substantial executability--acceptability gap: successfully executed simulations often fail to satisfy prescribed numerical or quantity-of-interest (QoI) criteria. Second, domain expertise in existing systems is tightly coupled with bespoke agent architectures, limiting its reuse and continual refinement across agent implementations. To address these challenges, we introduce ExtFoamBench, a challenging benchmark for OpenFOAM agents that
spans five axes of workflow variation, and separately assesses execution completion, numerical soundness, and QoI satisfaction. Complementing this, we propose EasyFoam, an agent framework that decouples CFD expertise into modular, externalized skills driven by a dual-loop mechanism. An execution loop constructs workflows from explicit task requirements and uses structured simulation evidence to diagnose failures and guide targeted repairs.
Meanwhile, an evolution loop distills task experience into reusable skill updates selected through paired evaluation. Empirical evaluations demonstrate that EasyFoam achieves substantially higher task acceptance rates than baseline agents, establishing this work as a starting point for future research toward self-improving CFD agents.
\end{abstract}

\section{Introduction}

Computational fluid dynamics (CFD) translates governing equations into a tightly coupled workflow spanning problem formulation, domain and mesh construction, physical and closure modeling, discretization, numerical solving, and quantity-of-interest (QoI) evaluation. The OpenFOAM ecosystem~\citep{weller1998tensorial} provides extensive flexibility across solvers, boundary conditions, turbulence closures, and post-processing utilities. This flexibility also makes autonomous CFD particularly challenging: an agent must coordinate interdependent decisions across multiple configuration files and execution stages while preserving consistency with the intended physical problem.

Recent advances in large language model (LLM) agents demonstrate substantial progress toward automating simulation pipelines. Systems such as MetaOpenFOAM~\citep{chen2024metaopenfoam}, OpenFOAMGPT~\citep{pandey2025openfoamgpt}, and Foam-Agent~\citep{yue2025foamagent} combine domain retrieval, multi-agent collaboration, and iterative syntax correction to construct and run OpenFOAM cases.
Benchmarks such as CFDLLMBench, which includes FoamBench, provide a basis for evaluating automated CFD workflows~\citep{somasekharan2025cfdllmbench}.
These developments demonstrate that modern LLM agents can increasingly automate end-to-end CFD workflows. However, two important limitations remain.

First, current evaluation paradigms are largely confined to simplistic, tutorial-derived tasks that modify scalar parameters or boundary conditions,   providing limited coverage of engineering-scale systems, coupled configurations, and multistage workflows. More fundamentally, successful execution does not guarantee an acceptable outcome. An OpenFOAM solver may terminate normally with apparently converged residuals despite inappropriate physical models, boundaries, or numerical settings. We define this systematic discrepancy as the \emph{executability--acceptability gap}: operationally executable simulations may fail essential numerical checks or task-specific QoI criteria.

Second, the CFD expertise in existing agents is often tightly coupled with specific prompts, agent roles, or workflow implementations. This monolithic design hinders knowledge reuse across different models or frameworks.
Crucially, it also limits continual improvement: successful workflow repairs typically remain isolated within local execution trajectories rather than being distilled into validated, reusable capabilities. Thus, achieving generalizable CFD automation demands a paradigm shift where domain knowledge is externalized, reused, and continuously evolved independently of the underlying agent architecture.

We address the first limitation by introducing \textbf{ExtFoamBench}, which extends reference-based CFD-agent evaluation to broader physical settings and multistage workflows. ExtFoamBench extends existing OpenFOAM benchmarks with broader physical systems, engineering-scale meshes, cross-stage configuration dependencies, and multistage execution protocols. To assess whether generated solutions meet task requirements beyond successful execution, ExtFoamBench evaluates them across three complementary dimensions: \emph{Execution Completion}, which verifies successful workflow execution; \emph{Numerical Soundness}, which assesses stability, boundedness, convergence, and related numerical requirements; and \emph{QoI Satisfaction}, which evaluates task-specific physical constraints and reference consistency. We report the proportion of tasks passing all three dimensions as the Task Acceptance Rate (TAR), explicitly exposing the executability--acceptability gap rather than hiding it behind solver termination.

To address the second limitation, we propose \textbf{EasyFoam}, an evidence-guided CFD agent framework driven by externalized CFD Skills and organized around a coupled dual-loop architecture. The \emph{execution loop} constructs and runs CFD workflows from explicit simulation requirements, monitors structured workflow state and evidence, and leverages numerical and QoI checks to diagnose failures, guiding targeted repair, re-execution, and reassessment.
The \emph{evolution loop} distills validated and regression-tested repairs into reusable skill updates, allowing CFD expertise to remain independent of the underlying agent implementation. These refined skills, in turn, enhance subsequent case construction, diagnosis, and repair. Thus, EasyFoam establishes a closed cycle of \emph{execution, evaluation, experience distillation, and skill evolution}, driving continual, evidence-grounded improvement.

Experiments on ExtFoamBench expose a substantial executability--acceptability gap: baseline agents can execute workflows successfully yet achieve near-zero TAR on complex cases, confirming that execution success can substantially overestimate actual problem-solving capability.
Under this evaluation, EasyFoam achieves the highest TAR among the evaluated systems on both FoamBench and ExtFoamBench, with skill evolution yielding further improvements on held-out tasks.
Crucially, while satisfying task-specific QoI criteria remains an open challenge, our findings validate a critical paradigm shift: transitioning CFD agents from mere workflow executors into autonomous systems that evaluate simulation evidence, perform targeted repairs, and continually evolve their problem-solving capabilities.

\section{ExtFoamBench}
\label{sec:ExtFoamBench}

ExtFoamBench evaluates whether CFD agents can construct complex simulation workflows and satisfy explicit numerical and scientific requirements. This section introduces its task construction and the multi-dimensional evaluation protocol used to distinguish successful execution from true task acceptance.

\subsection{Study-Level Task Construction}

Existing benchmarks provide a foundational evaluation of CFD agents but exhibit critical limitations in task diversity. Specifically, FoamBench-Basic primarily evaluates parameter sweeps within fixed, tutorial-derived configurations, while FoamBench-Advanced contains only 16 cases, with variation largely concentrated in turbulence models and geometry~\citep{somasekharan2025cfdllmbench}.
\begin{figure}[t]
    \centering
    \includegraphics[width=\linewidth]{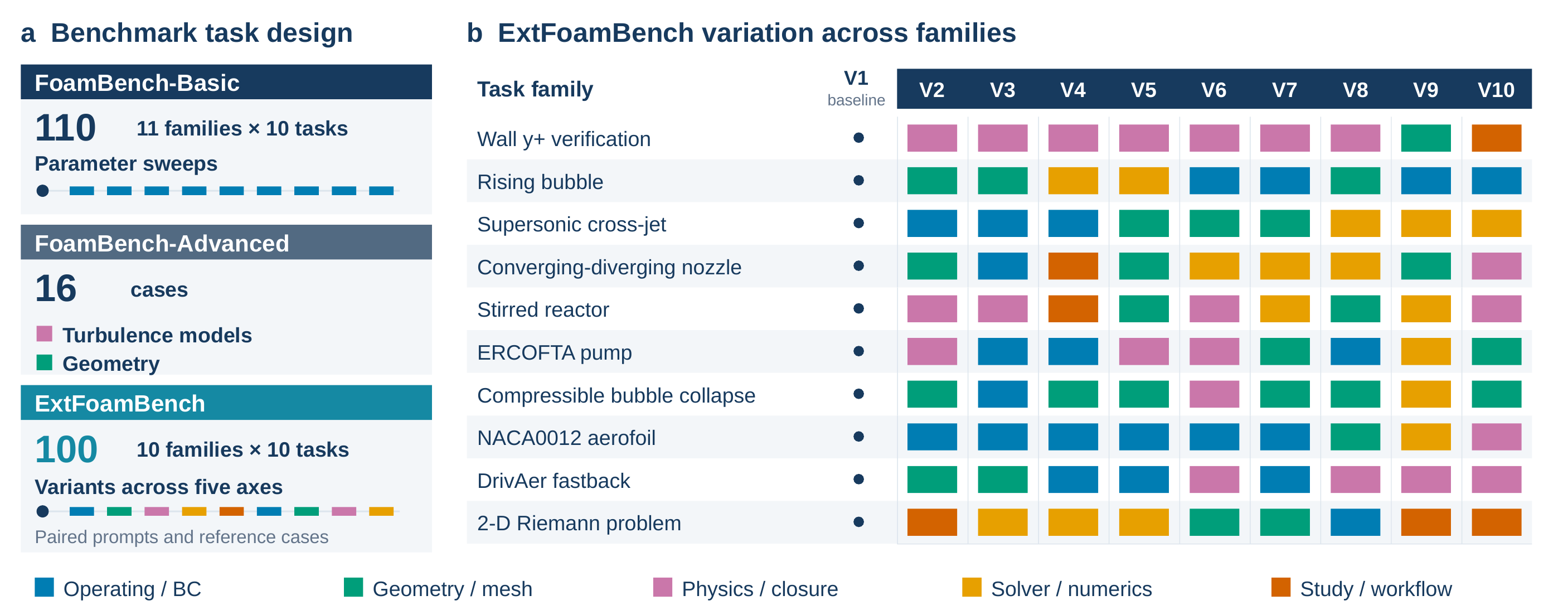}
    \caption{
    \textbf{Task construction and variation in FoamBench and ExtFoamBench.}
    \textbf{(a)} FoamBench-Basic primarily uses parameter sweeps, while FoamBench-Advanced contains 16 cases with variation concentrated in turbulence models and geometry. ExtFoamBench contains ten families, each with one baseline and nine variants.
    \textbf{(b)} Colors indicate the primary variation axis of each variant.
    }
    \label{fig:ExtFoamBench-design}
\end{figure}
ExtFoamBench broadens physical coverage and within-family workflow variation through 100 tasks across ten physical families (Figure~\ref{fig:ExtFoamBench-design}).

The ten families in ExtFoamBench are carefully curated to span a diverse spectrum of engineering-scale fluid dynamics, introducing distinct physical modeling and numerical stabilization challenges. Specifically, the tasks encompass \textbf{external aerodynamics} (NACA0012 aerofoil and DrivAer fastback), \textbf{compressibility and shock-wave interactions} (supersonic jet-in-crossflow, converging-diverging nozzle flow, and 2-D Riemann problems), \textbf{interface transport and two-phase dynamics} (rising bubbles and compressible bubble collapse), as well as \textbf{rotating-region modeling} (stirred reactors and rotating pumps). Additionally, a dedicated wall-$y^+$ assessment family explicitly evaluates near-wall resolution and turbulence treatment consistency. All baseline cases are adapted from established public sources to OpenFOAM Foundation v13 (see Appendix~\ref{app:ExtFoamBench-composition} for more details).

Each family contains one baseline configuration and nine purpose-built variants. Across the benchmark, these variants span five axes: operating and boundary conditions, geometry and mesh, physics and closure models, solver and numerical settings, and workflow execution protocols. Figure~\ref{fig:ExtFoamBench-design}(b) shows their distribution across families, with each variant labeled by its primary axis and potentially requiring accompanying changes along other axes. For example, changing a turbulence closure may require corresponding updates to transported fields and wall treatments, while modifying the mesh or geometry may affect boundary assignments and measurement locations. In the 2-D Riemann grid study, the agent must coordinate runs at three resolutions and extract and align solution profiles for comparison. ExtFoamBench therefore evaluates an agent's ability to adapt complete numerical workflows, requiring not only a runnable OpenFOAM case but also the specialized setup, execution procedures, and task-specific measurements and post-processing for task success.

\subsection{Multi-dimensional Scientific Evaluation}
\label{bench:evaluation}

Because successful solver termination does not guarantee an acceptable outcome, we evaluate ExtFoamBench and both FoamBench tracks using task-specific criteria. These criteria specify the required evidence and acceptance conditions across three complementary dimensions:

\begin{itemize}
\item \textbf{Execution Completion}: whether the declared preprocessing, solver execution, and required auxiliary utilities complete successfully;
\item \textbf{Numerical Soundness}: whether the computed solution satisfies task-specific numerical requirements, including stability, boundedness, convergence behavior, and mesh quality;
\item \textbf{QoI Satisfaction}: whether the resulting response satisfies task-specific physical constraints, reference relations, expected trends, or QoI tolerances.
\end{itemize}
For a given task \(\tau\), each evaluation dimension contains a set of mandatory criteria. Every criterion is recorded as \texttt{pass}, \texttt{fail}, or \texttt{unknown}: pass when the available evidence satisfies the criterion, fail when the evidence demonstrates a violation, and unknown when the evidence is missing or insufficient.
Let $E_\tau$, $N_\tau$, and $Q_\tau \in \{0, 1\}$ denote the binary pass indicators for these three dimensions, respectively. A dimension passes (indicator equals $1$) if and only if all of its mandatory criteria \(\mathcal{B}_{\tau,X} \) evaluate to \texttt{pass}: $X_\tau=\prod_{b\in\mathcal{B}_{\tau,X}}\mathbf{1}[b=\mathrm{pass}],\;X\in\{E,N,Q\}$.

The Execution Completion Rate (ECR) measures the proportion of tasks whose prescribed workflows complete successfully. However, a task is considered genuinely accepted if and only if all three evaluation dimensions pass simultaneously. Accordingly, over a task set \(\mathcal{D}\), we define ECR and the Task Acceptance Rate (TAR):
$$
\mathrm{ECR}(\mathcal{D}) = \frac{1}{|\mathcal{D}|}\sum_{\tau\in\mathcal{D}} E_\tau, \qquad \mathrm{TAR}(\mathcal{D}) = \frac{1}{|\mathcal{D}|}\sum_{\tau\in\mathcal{D}} E_\tau N_\tau Q_\tau.
$$
We adopt TAR as our primary performance metric. The difference between $\mathrm{ECR}(\mathcal{D})$ and $\mathrm{TAR}(\mathcal{D})$ formally quantifies the \emph{executability--acceptability gap}. Importantly, satisfying all criteria establishes a necessary, auditable verification baseline for simulation reliability, rather than an absolute guarantee of physical correctness.

\section{EasyFoam}
\label{sec:method}

\begin{figure}[t]
    \centering
    \includegraphics[width=\linewidth]{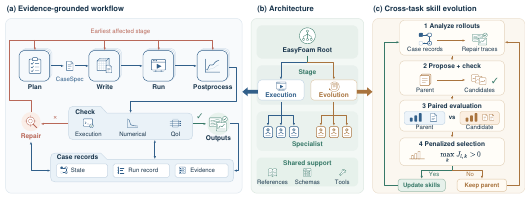}
    \caption{\textbf{Overview of EasyFoam.}
    \textbf{(a)} Execution loop: evidence-guided construction and stage-scoped repair.
    \textbf{(b)} Architecture: hierarchical skill organization.
    \textbf{(c)} Evolution loop: experience distillation via paired evaluation.
    }
    \label{fig:method}
\end{figure}

EasyFoam externalizes CFD expertise into reusable skills and connects their application and refinement through a coupled dual-loop mechanism (Figure~\ref{fig:method}). Within a task, the execution loop uses structured simulation evidence to construct, assess, and repair the current workflow. Across tasks, the evolution loop converts execution experience into candidate skill revisions and selects updates through paired evaluation with regression penalties. Together, they form a mutually reinforcing cycle: the execution loop generates the practical experience needed for refinement, while the evolved skills, in turn, improve case construction and repair in subsequent tasks.

\subsection{Externalized CFD Skill Architecture}
\label{subsec:skill_arthitecture}

EasyFoam encapsulates CFD knowledge as reusable skills that guide the underlying coding agent in case construction, diagnosis, and repair. To effectively separate open-ended CFD reasoning from reproducible operations, the capability system relies on four integrated pillars: (1) \textbf{Hierarchical Skills} that define execution boundaries and decision procedures; (2) \textbf{References} providing read-only domain knowledge; (3) \textbf{Schemas} defining canonical workflow artifacts; and (4) \textbf{Tools} containing deterministic operations for state management and evidence extraction.

As shown in Figure~\ref{fig:method}(b), EasyFoam organizes Skills in a Root--Stage--Specialist hierarchy. The \emph{Root} routes task intent to either the Execution branch or the Evolution branch; the \emph{Stage} defines stage entry, completion and handoff conditions; and the \emph{Specialist} guides bounded activities such as case planning, meshing, case authoring, runtime diagnosis, and post-processing.
When the agent loads the skills relevant to the current decision, shared case records connect these procedures, allowing downstream stages to reliably access the artifacts produced upstream.

\subsection{In-Task Execution Loop}
\label{subsec:execution-loop}

The execution loop uses the active skill system to translate user requirements into a functional CFD workflow, iteratively assessing and repairing the case based on runtime evidence. As illustrated in Figure~\ref{fig:method}(a), it orchestrates three core operations.

\paragraph{Task specification and evidence planning.}
CFD requests involve interdependent physical conditions, numerical settings, and output requirements. When these are scattered, agents may omit essential settings or produce insufficient evidence for final assessment. To address this, the Plan stage consolidates these requirements into a shared task specification, \texttt{CaseSpec}, covering the physical setup (models, properties, boundaries), computational setup (geometry, numerical schemes), and requested QoIs and extraction requirements. Provenance tracks whether a decision is an explicit requirement, a domain-based derivation, a permitted default, or an unresolved assumption. Crucially, any unconfirmed assumptions affecting the physical problem or target outcomes are explicitly marked as unresolved, blocking execution until clarified.

\paragraph{Workflow construction and inspectable state.}
The workflow proceeds sequentially: \textbf{Plan} $\rightarrow$ \textbf{Write} $\rightarrow$ \textbf{Run} $\rightarrow$ \textbf{Postprocess}.
\textbf{Plan} defines the task specification and execution strategy; \textbf{Write} prepares the physical models, mesh, fields, and numerical configurations; \textbf{Run} executes the prescribed solvers and auxiliary utilities; and \textbf{Postprocess} extracts the requested QoIs and assessment evidence. Each stage uses upstream artifacts and follows the same task specification. For instance, the generated mesh inventory inherently constrains subsequent boundary definitions, and the executed temporal coverage dictates which QoIs can be successfully extracted. Workflow progression therefore strictly requires completed operations and sufficient valid outputs to support the subsequent stage, establishing a clear dependency chain that lays the groundwork for precise error localization.

\paragraph{Evidence-guided stage-scoped repair.}
Relying on the multi-dimensional criteria defined in Section~\ref{bench:evaluation}, EasyFoam systematically audits the workflow states and outcomes at multiple levels. During execution, these checks use task requirements and available simulation evidence; hidden reference-dependent criteria are reserved for offline evaluation and are never exposed to the solving agent. A failed or unknown mandatory criterion immediately triggers a diagnosis to isolate the earliest implicated workflow stage. For example, a missing QoI returns to Postprocess if the fields exist, whereas a fundamental configuration defect returns the agent to Write. The relevant Specialist then guides the targeted repair. Crucially, after repairing the implicated stage, EasyFoam automatically invalidates and regenerates all dependent downstream results. This \emph{stage-scoped repair} preserves valid upstream work while strictly preventing stale evidence from falsely validating the repaired case. The loop terminates when all mandatory checks pass or the execution budget is exhausted.

\subsection{Experience-Driven Skill Evolution}
\label{subsec:skill-evolution}

Resolving a single case is insufficient for generalizable automation; the benefits remain local unless translated into persistent capabilities. EasyFoam converts in-task repair experience into candidate skill updates and selects them through paired evaluation on separate validation tasks (Figure~\ref{fig:method}(c)).

\paragraph{Turning experience into skill updates.}
At round $l$, a proposal agent examines diagnostic feedback and execution traces from a designated set of construction tasks. By analyzing these trajectories alongside the current skill instructions $\mathcal{K}_l$, the agent identifies underlying procedural flaws and proposes updates designed to generalize across tasks. For instance, observing repeated field--patch mismatches might prompt the agent to explicitly encode a mesh-inventory verification step within the boundary-authoring skill.
By generating multiple candidate skill revisions $\mathcal{K}_l^{(k)}$ sequentially, the system systematically explores diverse fixes for both numerical and evidence-collection bottlenecks. These candidates then undergo preliminary structural checks to filter out invalid or redundant updates before advancing to formal held-out validation.

\paragraph{Measuring progress toward task acceptance.}
Task acceptance requires all three evaluation dimensions to pass jointly. However, on highly complex tasks, this strict binary outcome can conceal valuable intermediate progress (e.g., a revision that improves numerical behavior but does not yet satisfy the requested QoI). To provide denser feedback, EasyFoam guides evolution using a reward function that credits passes in individual evaluation dimensions:
\begin{equation}
r_\tau(\mathcal{K}) = \alpha_E E_\tau(\mathcal{K}) + \alpha_N N_\tau(\mathcal{K}) + \alpha_Q Q_\tau(\mathcal{K}).
\label{eq:reward}
\end{equation}
The fixed, nonnegative weights $\alpha$ sum to one, with the highest weight assigned to QoI Satisfaction to emphasize task-specific physical requirements.

\paragraph{Selecting updates through paired evaluation.}
To prevent overfitting, the parent skill system and eligible candidates are rigorously evaluated on a disjoint set of validation tasks under strictly matched backends and budgets. Let $R_{l,d}(\mathcal{K}_l)$ denote the mean task reward achieved by the parent system on task family $d$, and let $R_{l,d}(\mathcal{K}_l^{(k)})$ denote the reward of the $k$-th candidate. The performance gain of candidate $k$ is $\Delta_{l,d}^{(k)} = R_{l,d}(\mathcal{K}_l^{(k)}) - R_{l,d}(\mathcal{K}_l)$. EasyFoam computes the final selection objective $J_{l,k}$ by combining these gains with a family-level regression penalty:
\begin{equation}
J_{l,k} = \sum_d w_{l,d}\Delta_{l,d}^{(k)} - \lambda \sum_d w_{l,d}\max(0, -\Delta_{l,d}^{(k)}),
\label{eq:penalty}
\end{equation}
where $w_{l,d} \geq 0$ are the weights for each family ($\sum_d w_{l,d} = 1$), and $\lambda > 0$ controls the severity of the penalty for negative gains. The candidate with the largest strictly positive $J_{l,k}$ replaces the parent to become $\mathcal{K}_{l+1}$; if no candidate improves the objective, the parent is retained. Subsequent rounds build upon the retained version, allowing skill refinement to accumulate across rounds.

\section{Experiments}
\label{sec:experiments}

\subsection{Experimental Setup}
\label{sec:exp-setup}

We evaluate EasyFoam on FoamBench (126 tasks) and ExtFoamBench (100 tasks), comparing it with three CFD agent baselines: MetaOpenFOAM 2.0~\citep{chen2025metaopenfoam}, Foam-Agent~\citep{yue2025foamagent}, and ChatCFD~\citep{fan2025chatcfd}. All systems use DeepSeek-V4-Flash~\citep{deepseekai2026deepseekv4} as the LLM backbone and OpenFOAM Foundation v13. System-specific runtime settings are provided in Appendix~\ref{app:runtime-settings}. The EasyFoam results in Table~\ref{tab:main-results} use the manually authored initial Skills, before any cross-task skill evolution.

\subsection{In-task Performance}
\label{sec:acceptance-performance}

\begin{table}[t]
\caption{System-level comparison. Entries report passed tasks over all assigned tasks and percentages; TAR requires joint passage of the three evaluation dimensions.}
\label{tab:main-results}
\centering
\small
\resizebox{\linewidth}{!}{%
\begin{tabular}{llcccc}
\toprule
Track & Method & \shortstack{Execution\\Completion} & \shortstack{Numerical\\Soundness} & \shortstack{QoI\\Satisfaction} & TAR \\
\midrule
\multirow{4}{*}{FoamBench (126)}
& MetaOpenFOAM 2.0 & 36/126 (28.6\%) & 29/126 (23.0\%) & 7/126 (5.6\%) & 6/126 (4.8\%) \\
& Foam-Agent & 75/126 (59.5\%) & 46/126 (36.5\%) & 5/126 (4.0\%) & 5/126 (4.0\%) \\
& ChatCFD & 44/126 (34.9\%) & 41/126 (32.5\%) & 21/126 (16.7\%) & 21/126 (16.7\%) \\
& EasyFoam & \textbf{123/126 (97.6\%)} & \textbf{94/126 (74.6\%)} & \textbf{55/126 (43.7\%)} & \textbf{52/126 (41.3\%)} \\
\midrule
\multirow{4}{*}{ExtFoamBench (100)}
& MetaOpenFOAM 2.0 & 8/100 (8.0\%) & 6/100 (6.0\%) & 0/100 (0.0\%) & 0/100 (0.0\%) \\
& Foam-Agent & 27/100 (27.0\%) & 21/100 (21.0\%) & 0/100 (0.0\%) & 0/100 (0.0\%) \\
& ChatCFD & 13/100 (13.0\%) & 11/100 (11.0\%) & 0/100 (0.0\%) & 0/100 (0.0\%) \\
& EasyFoam & \textbf{72/100 (72.0\%)} & \textbf{62/100 (62.0\%)} & \textbf{7/100 (7.0\%)} & \textbf{6/100 (6.0\%)} \\
\bottomrule
\end{tabular}
}
\end{table}

\textbf{Execution versus acceptance.}
EasyFoam leads across all four metrics on FoamBench (Table~\ref{tab:main-results}). The baseline comparison illustrates the executability--acceptability gap: Foam-Agent completes more workflows than ChatCFD, yet ChatCFD achieves a higher TAR. This performance reversal demonstrates that successful execution does not reliably distinguish systems capable of satisfying numerical and task-specific QoI requirements.

\textbf{QoI satisfaction on ExtFoamBench.}
All systems achieve substantially lower execution completion rates on ExtFoamBench than on FoamBench, reflecting greater difficulty in constructing and completing its workflows. Despite this difficulty, EasyFoam passes QoI checks on seven tasks, with six satisfying all acceptance criteria, whereas none of the three baselines passes the QoI evaluation. Task-level diagnosis reveals that among the 56 tasks passing numerical checks but failing QoI evaluation, 51 extracted the required metrics but violated at least one physical reference-comparison criterion (e.g., incorrect bubble geometry, near-wall $y^+$, or flow-rate quantities). The challenge thus extends beyond completing the workflow and obtaining numerically acceptable solutions to producing responses that meet task-specific requirements.

\begin{figure}[t]
\centering
\includegraphics[width=\linewidth]{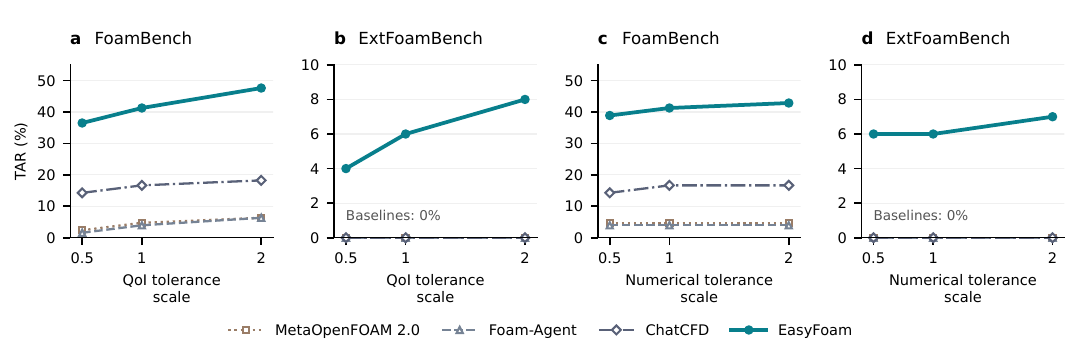}
\caption{TAR sensitivity to QoI tolerances (a,b) and Numerical tolerances (c,d) on FoamBench (a,c) and ExtFoamBench (b,d).  The three baselines in (b,d) overlap at zero.}
\label{fig:threshold-sensitivity}
\end{figure}

\textbf{Sensitivity to acceptance tolerances.}
Reference-based acceptance checks allow a specified amount of error: for example, a predicted drag coefficient passes if its deviation from the reference falls within the prescribed tolerance. We examine how these allowed errors affect TAR by scaling the Numerical or QoI tolerances to $0.5$, $1$, and $2$ times their original values, while keeping the other group unchanged. The same simulation outputs are then reassessed under these stricter or looser criteria. Figure~\ref{fig:threshold-sensitivity} shows that EasyFoam consistently outperforms the baselines on both benchmarks, achieving higher TAR under stricter tolerances than all baselines under relaxed tolerances. Even when tolerances are doubled, the baselines show only modest gains on FoamBench and remain at zero TAR on ExtFoamBench, suggesting persistent limitations in jointly satisfying numerical and QoI requirements. Jointly relaxing both groups yields the same overall pattern (Appendix~\ref{app:threshold-sensitivity}). The limited recovery under relaxed tolerances suggests that substantial progress will require improving the generated solutions themselves, beyond changing where the acceptance boundary is drawn.

\subsection{Ablation Study}
\label{sec:component-analysis}

Due to API cost constraints, we conduct ablations on a subset of FoamBench (63 tasks), replacing stage-scoped repair with whole-case retry or progressively removing repair and skills from the full EasyFoam system. The \textbf{w/o Stage-Scoped Repair} variant retains domain skills, workflow guidance, and task-visible checks, but replaces stage-scoped repair with whole-case retry. The \textbf{w/o Repair} variant retains domain skills, workflow guidance, and task-visible checks, but disables repair. The \textbf{w/o Repair \& Skills} variant further removes the skills and associated workflow guidance, serving as our generic coding-agent baseline. Appendix~\ref{app:ablation-protocol} details the ablation settings.

\begin{table}[t]
\caption{Ablation of EasyFoam on 63 FoamBench tasks. Entries report passed tasks and percentages.}
\label{tab:component-ablation}
\centering
\small
\resizebox{\linewidth}{!}{%
\begin{tabular}{lcccc}
\toprule
Method Variant & \shortstack{Execution\\Completion} & \shortstack{Numerical\\Soundness} & \shortstack{QoI\\Satisfaction} & TAR \\
\midrule
EasyFoam & \textbf{63/63 (100.0\%)} & \textbf{53/63 (84.1\%)} & \textbf{37/63 (58.7\%)} & \textbf{35/63 (55.6\%)} \\
w/o Stage-Scoped Repair & 57/63 (90.5\%) & 46/63 (73.0\%) & 33/63 (52.4\%) & 32/63 (50.8\%) \\
w/o Repair & 51/63 (81.0\%) & 43/63 (68.3\%) & 32/63 (50.8\%) & 31/63 (49.2\%) \\
w/o Repair \& Skills & 49/63 (77.8\%) & 44/63 (69.8\%) & 23/63 (36.5\%) & 23/63 (36.5\%) \\
\bottomrule
\end{tabular}
}
\end{table}

The ablation results suggest that skills, repair, and the recovery strategy contribute to task acceptance (Table~\ref{tab:component-ablation}). Execution traces from \texttt{shallowWaterWithSquareBump/1} illustrate how the recovery strategy affects the correction of an identified defect, while \texttt{squareBend/4} illustrates the complementary roles of skills and repair. During case construction, skills guide the agent in translating task requirements into settings across multiple configuration files. Both the w/o Repair and full EasyFoam rollouts specify the required flow conditions and fluid properties, together with the requested run endpoint and output interval (Appendix~\ref{app:repair-evidence}).

However, specifying these requirements in the configuration did not ensure that they were met during execution. In both independent rollouts, the solver exited normally after 145 steady-state iterations: a convergence-based stopping rule inherited from the tutorial terminated the run before the prescribed endpoint of 500, leaving the requested output schedule incomplete. The w/o Repair variant identified the conflicting stopping rule but did not modify the case. Full EasyFoam removed this rule (the \texttt{residualControl} block in \texttt{fvSolution}), revalidated the configuration, and reran the solver without changing the mesh or physical boundary conditions. The revised case reached the prescribed endpoint and passed all acceptance criteria (Appendix~\ref{app:repair-evidence}). This example illustrates how repair complements skills-guided construction by resolving a conflict between the prescribed endpoint and the solver's stopping behavior. The comparison with w/o Stage-Scoped Repair further illustrates the importance of acting on the diagnosed defect. In \texttt{shallowWaterWithSquareBump/1}, both variants reached the prescribed 100\,s endpoint and identified that indented solver time lines prevented completion verification. The w/o Stage-Scoped Repair variant repeated the complete execution chain without correcting the detection mismatch, reproducing the blocked run record. Full EasyFoam normalized the solver output through a Run-stage wrapper before re-execution, restoring verifiable completion without changing the recorded case inputs or mesh (Appendix~\ref{app:stage-scoped-repair-evidence}).

\subsection{Cross-Task Skill Evolution}
\label{sec:skill-evolution-experiment}

We evaluate the initial and final evolved skills on a strictly disjoint set of 57 held-out test tasks. These tasks are excluded from candidate generation and selection. Appendix~\ref{app:evolution-experiment} specifies the evolution procedure and data partitioning.

\begin{table}[t]
\caption{Initial and evolved skills on the same 57 held-out tasks, with a frozen backbone and matched execution budgets.}
\label{tab:evolution-frozen-results}
\centering
\small
\resizebox{\linewidth}{!}{%
\begin{tabular}{llcccc}
\toprule
Test Track & Skill Library & \shortstack{Execution\\Completion} & \shortstack{Numerical\\Soundness} & \shortstack{QoI\\Satisfaction} & TAR \\
\midrule
\multirow{2}{*}{FoamBench (32)}
& Initial & 30/32 (93.8\%) & 23/32 (71.9\%) & 20/32 (62.5\%) & 16/32 (50.0\%) \\
& Evolved & \textbf{32/32 (100.0\%)} & \textbf{32/32 (100.0\%)} & \textbf{22/32 (68.8\%)} & \textbf{22/32 (68.8\%)} \\
\midrule
\multirow{2}{*}{ExtFoamBench (25)}
& Initial & 15/25 (60.0\%) & 10/25 (40.0\%) & 2/25 (8.0\%) & 2/25 (8.0\%) \\
& Evolved & \textbf{22/25 (88.0\%)} & \textbf{18/25 (72.0\%)} & \textbf{5/25 (20.0\%)} & \textbf{5/25 (20.0\%)} \\
\midrule
\multirow{2}{*}{Overall (57)}
& Initial & 45/57 (78.9\%) & 33/57 (57.9\%) & 22/57 (38.6\%) & 18/57 (31.6\%) \\
& Evolved & \textbf{54/57 (94.7\%)} & \textbf{50/57 (87.7\%)} & \textbf{27/57 (47.4\%)} & \textbf{27/57 (47.4\%)} \\
\bottomrule
\end{tabular}
}
\end{table}

The evolved skills improve TAR on both benchmarks with model weights fixed (Table~\ref{tab:evolution-frozen-results}). Gains are largest in Numerical Soundness, indicating that the evolution process effectively distills localized task experience into reusable guidance for robust diagnosis and execution. The retained skill updates provide explicit evidence of this capability transfer. For example, a nozzle failure involving nonphysical negative temperatures motivated an ordered procedure for checking boundary conditions, time-step behavior, and thermodynamic consistency before changing solver controls. Another update moved diagnostic-output preparation into case construction and required checking output availability before formal execution. In a subsequent shallow-water task, the agent applied this instruction by reading the required outputs from a short trial run, establishing that the planned analysis was supported before committing to the full simulation. These updates preserve procedures for investigating failures and preparing usable outputs beyond the cases that motivated them.

The updates also make sampling adequacy an explicit requirement for interpreting simulation outputs, directing the agent to check the recorded interval and frequency and obtain additional data when needed. In \texttt{BernardCells/10}, an apparent heat-flux imbalance prompted the evolved agent to inspect the underlying records. Heat flux was recorded every 50 seconds, whereas temperature was recorded every second. Following the updated guidance, the agent resumed from a saved state and generated one-second heat-flux records during a diagnostic run. Our reanalysis over the same interval yielded an imbalance of $-0.638\%$ using all samples, compared with $-1.634\%$ using only every fiftieth sample. Sparse sampling thus exaggerated the apparent imbalance in this case. The evolved procedure made sampling adequacy part of the diagnosis, providing a better basis for deciding whether the discrepancy warranted changes to the simulation. Appendix~\ref{app:evolution-evidence} provides the instruction changes, execution traces, and sampling comparison.

\section{Related Work}
\label{sec:related_work}

\paragraph{LLM Agents and Benchmarks for CFD.}
Recent systems automate OpenFOAM workflows through diverse domain guidance strategies. For instance, MetaOpenFOAM decomposes tasks via role-based multi-agent collaboration and tutorial retrieval~\citep{chen2024metaopenfoam}, whereas OpenFOAMGPT relies on retrieval-augmented generation~\citep{pandey2025openfoamgpt}.
Workflow capabilities are further extended by Foam-Agent, which utilizes hierarchical retrieval and composable tools for end-to-end meshing and visualization~\citep{yue2025foamagent}, and ChatCFD, which emphasizes structured reasoning and error reflection~\citep{fan2025chatcfd}. To evaluate these systems, CFDLLMBench and its FoamBench component assess workflow implementations against tutorial and expert-designed cases~\citep{somasekharan2025cfdllmbench}.
ExtFoamBench builds upon this foundation by systematically broadening physical and workflow variations, shifting the evaluation towards a strict separation of execution completion, numerical soundness, and task-specific QoI satisfaction.

\paragraph{Evidence-Guided Correction in Scientific Simulation.}
Interactive agents iteratively refine solutions using execution feedback and reflection~\citep{yao2023react,shinn2023reflexion,madaan2023selfrefine}. In scientific simulation, \emph{silent failures} can persist despite normal solver termination or apparently converged residuals, motivating verification and self-correction~\citep{park2026self}. In CFD, PhyNiKCE~\citep{fan2026phynikce} applies symbolic constraints to simulation setup, while AI CFD Scientist~\citep{somasekharan2026ai} uses rendered flow fields via vision-language models for physics verification. EasyFoam advances this paradigm by explicitly linking task-visible numerical and QoI checks to specific workflow states. By treating evaluation outcomes as structured evidence, EasyFoam drives stage-scoped repairs and invalidates dependent downstream artifacts, directly connecting scientific assessment to the workflow stages implicated by the failure.

\paragraph{Reusable Agent Skills and Skill Evolution.}
Agents increasingly leverage external libraries to accumulate procedural knowledge, ranging from interactive skill building~\citep{wang2023voyager} and evolving context playbooks~\citep{zhang2026agentic} to trajectory-based prompt optimization~\citep{agrawal2026gepa}. Frameworks like EvoSkill and SkillOpt directly refine these skills via execution feedback~\citep{alzubi2026evoskill, yang2026skillopt}, while recent reinforcement-learning approaches optimize skill generators through reward-guided evolution~\citep{xia2026skillrl, vishe2026skill}. While these methods demonstrate the power of experience-driven adaptation, execution-based rewards are insufficient in CFD when solver completion is used as the success signal, because completion does not imply satisfaction of task-specific numerical or QoI criteria. EasyFoam translates localized task experience into reusable CFD skill revisions and selects updates through paired evaluation that accounts for execution, numerical, and QoI outcomes while penalizing family-level regressions.

\section{Conclusion and Limitations}
We introduce ExtFoamBench to strictly evaluate agents across execution completion, numerical soundness, and QoI satisfaction. Complementing this, we introduce EasyFoam, a framework that externalizes CFD expertise as reusable skills, uses simulation evidence to guide stage-scoped repair within tasks, and improves these skills across tasks through paired evaluation. Our experiments demonstrate that EasyFoam significantly outperforms existing baselines in overall TAR. Crucially, the systemic discrepancy between execution completion and TAR across all systems reveals the critical executability--acceptability gap in automated CFD workflows. While EasyFoam successfully distills diagnostic experience to improve numerical reliability on held-out tasks, satisfying complex physical constraints in more complex and out-of-family regimes remains a formidable frontier.

Our study has several limitations. ExtFoamBench does not yet cover strongly coupled problems such as two-way interaction between fluids and deformable structures, or multi-region conjugate heat transfer with phase change. These settings require coordinating additional physical models and maintaining consistent exchanges across coupled interfaces. Our implementation and experiments also focus on OpenFOAM, leaving adaptation to other CFD software packages for future work.

\section*{AI Use Statement}
Generative AI tools assisted with manuscript organization, language refinement, and presentation feedback. The authors reviewed and verified the benchmark specifications, CFD terminology, evaluation criteria, mathematical formulations, experimental results, and final manuscript content. The use of LLMs within EasyFoam and the evaluated baseline agents is described separately in the methodology and experimental settings. The authors take full responsibility for the accuracy and integrity of the work.

\section*{Reproducibility Statement}
Appendices A--E provide benchmark construction details, evaluation criteria, implementation details, experimental settings, case studies, and resource usage.

\bibliography{references}
\bibliographystyle{iclr2027_conference}

\clearpage
\appendix

\section{Benchmark Construction}
\label{app:ExtFoamBench}

\subsection{Sources and Task Variants}
\label{app:ExtFoamBench-composition}

ExtFoamBench contains ten physical families, each with one baseline and nine variants. Table~\ref{tab:ExtFoamBench-families} lists their sources and adapted configurations, while Table~\ref{tab:variants-one} details the within-family variations. Seven families draw on dyfluid examples; NACA0012 and DrivAer fastback use OpenFOAM tutorial cases and geometry assets~\citep{greenshields2025}; and the two-dimensional Riemann family draws on a CFD-China four-quadrant problem. The nozzle and bubble-collapse cases additionally draw on the physical backgrounds described by \citet{slater2001cdv} and \citet{beig2018temperatures}, respectively. All cases are adapted to OpenFOAM Foundation v13.

\begin{table}[htbp]
\centering\footnotesize
\renewcommand{\arraystretch}{1.12}
\caption{ExtFoamBench family sources and OpenFOAM v13 configurations; ten tasks per family.}
\label{tab:ExtFoamBench-families}
\begin{tabularx}{\linewidth}{@{}>{\raggedright\arraybackslash}p{0.18\linewidth}>{\raggedright\arraybackslash}p{0.23\linewidth}Y@{}}
\toprule
Family & Starting source & Base configuration \\
\midrule
Wall $y^+$ & dyfluid channel example & Periodic turbulent half-channel with mean-velocity forcing and wall-normal sampling. \\
\midrule
Rising bubble & dyfluid VOF example & Two-phase rising bubble with prescribed phase properties and initial interface. \\
\midrule
Supersonic jet-in-crossflow & dyfluid compressible-flow example & Quasi-2D supersonic slot jet with prescribed inlet thermodynamic states. \\
\midrule
CDV nozzle & dyfluid nozzle example; NASA CDV background & Compressible flow through a supplied converging--diverging nozzle. \\
\midrule
Stirred reactor & dyfluid MRF example; OpenFOAM Rushton assets & MRF stirred vessel with supplied impeller and baffle surfaces. \\
\midrule
ERCOFTA pump & dyfluid rotating-machinery example & Rectangular rotating-region surrogate with specified rotor zone, inlet and outlet. \\
\midrule
Compressible bubble collapse & dyfluid compressible-VOF example; JFM background & Compressible two-phase collapse in an axisymmetric wedge with specified equations of state. \\
\midrule
NACA0012 & OpenFOAM tutorial and surface asset & Quasi-2D compressible aerofoil flow with prescribed force normalization. \\
\midrule
DrivAer fastback & OpenFOAM tutorial and body/wheel assets & External aerodynamics around supplied body and wheel surfaces. \\
\midrule
2-D Riemann & CFD-China four-quadrant problem & Four-quadrant compressible flow with prescribed initial states and centre-line sampling. \\
\bottomrule
\end{tabularx}
\end{table}

\begin{table}[htbp]
\centering\footnotesize
\renewcommand{\arraystretch}{1.15}
\caption{ExtFoamBench variation design across all ten physical families. V1 denotes the baseline; V2--V10 identify variants.}
\label{tab:variants-one}
\begin{tabularx}{\linewidth}{@{}>{\raggedright\arraybackslash}p{0.18\linewidth}Y@{}}
\toprule Family & Changes in V2--V10 \\
\midrule
Wall $y^+$ & V2: $k$--$\epsilon$; V3: Launder--Sharma; V4: Lam--Bremhorst; V5: Lien cubic; V6: v2f; V7: $q$--$\zeta$; V8: $k$--$\omega$ SST; V9: wall-normal mesh grading/refinement; V10: viscosity comparison. V3--V7 also use finer wall-normal meshes and longer runs. \\
\midrule
Rising bubble & V2/V3: successive mesh refinement; V4: explicit interface compression; V5: PLIC interface fluxes with compression and MULES correction; V6/V7: reduced/increased surface tension; V8: interface AMR; V9: further reduced surface tension; V10: increased viscosities under the V9 condition. \\
\midrule
Supersonic jet-in-crossflow & V2: changed main-flow Mach number; V3/V4: lower/higher jet pressure ratio; V5/V6: changed injection angles; V7: streamwise refinement of the first active block; V8: unbounded upwind; V9: earlier switch to bounded limitedLinear; V10: transient \texttt{fluid} solver. \\
\midrule
CDV nozzle & V2: converging-section refinement; V3: changed back pressure; V4: multi-back-pressure study; V5: \texttt{polyDualMesh} conversion; V6: upwind-to-limitedLinear continuation; V7: total-pressure/total-temperature inlet; V8: \texttt{shockFluid}; V9: axisymmetric wedge; V10: steady $k$--$\omega$ SST RANS. \\
\midrule
Stirred reactor & V2: $k$--$\epsilon$; V3: $k$--$\omega$ SST; V4: rotation-rate comparison; V5: baffled/unbaffled comparison; V6: CrossPowerLaw rheology; V7: MRF/sliding-mesh NCC comparison; V8: doubled mesh factor; V9: transient PIMPLE; V10: water/oil VOF in the specified rotor--stator vessel. \\
\midrule
ERCOFTA pump & V2: compressible \texttt{fluid}; V3: off-design rotation; V4: increased outlet pressure; V5: transient $k$--$\omega$ SST DES; V6: $k$--$\epsilon$; V7: extended rotor zone; V8: reversed rotation; V9: transient PIMPLE; V10: uniform mesh refinement. \\
\midrule
Compressible bubble collapse & V2: initial-radius comparison; V3: ambient-pressure comparison; V4: near-wall placement; V5: stepped asymmetric initial contour; V6: lower gas density; V7: interface AMR; V8: uniform mesh refinement; V9: incompressible VOF comparison; V10: two bubbles. \\
\midrule
NACA0012 & V2--V4: increasing Mach number at zero incidence; V5--V7: negative and positive angles of attack at fixed Mach number; V8: multilevel \texttt{snappyHexMesh} refinement; V9: transient PIMPLE; V10: steady Spalart--Allmaras RANS. \\
\midrule
DrivAer fastback & V2/V3: streamwise mesh refinement; V4/V5: increased/reduced freestream speed; V6: MRF wheel treatment; V7: moving ground; V8: $k$--$\epsilon$; V9: $k$--$\omega$; V10: transient Spalart--Allmaras DDES. V2 and V8 additionally prescribe continuation and merging of force histories. \\
\midrule
2-D Riemann & V2: three-level grid study with successive resolution doubling; V3: Courant-limit comparison; V4: Kurganov/Tadmor comparison; V5: vanAlbada/vanLeer limiters; V6: rotated initial discontinuities and velocities; V7: domain-size comparison at fixed spacing; V8: strengthened pressure/density contrast; V9: serial/parallel decomposition comparison; V10: continuous/restarted calculation. \\
\bottomrule
\end{tabularx}
\end{table}

\subsection{Prompt--Reference Pairing}
\label{app:ExtFoamBench-workflow}

Each task pairs a manually written prompt with an independently executed reference case. The prompt specifies the required physical conditions, computational constraints, and requested outputs, while the reference case provides a corresponding implementation and the quantities used for offline evaluation. Required external geometry assets are supplied to the solving agent with source attribution; reference case configurations and computed reference QoIs remain hidden during the solving rollout.

For multi-run studies, the task specifies the required branches and their associated outputs. Branch identities are preserved throughout execution and evaluation so that quantities from different operating conditions, resolutions, or execution protocols can be associated with the appropriate comparison.

\section{Evaluation Details}
\label{app:evaluation-protocol}

\subsection{Execution Completion and Numerical Soundness Criteria}
\label{app:evaluation-numerical}

The offline evaluator assesses frozen case files, execution logs, solution fields, and derived outputs using task-specific criteria fixed before evaluation and applied consistently across systems. Execution Completion requires the prescribed preprocessing, solver runs, and auxiliary utilities to complete successfully within the rollout budget. Numerical Soundness and QoI Satisfaction are evaluated only for successfully executed tasks; incomplete execution receives no pass credit in either dimension.

Within each dimension, all mandatory criteria must be satisfied. A criterion is marked \textit{pass} when the available evidence satisfies its acceptance rule, \textit{fail} when the evidence demonstrates a violation, and \textit{unknown} when the required evidence is unavailable or insufficient. Both \textit{fail} and \textit{unknown} prevent the corresponding dimension from passing. Table~\ref{tab:numerical-checks} summarizes the Numerical Soundness criteria.

\begin{table}[htbp]
\centering\footnotesize
\renewcommand{\arraystretch}{1.12}
\caption{Numerical Soundness criteria. $x^\star$ denotes the corresponding reference diagnostic, and \(n\) is the number of available samples in the diagnostic series. Tail p95 is the 95th percentile over the final $\min(n,\max(5,\lceil0.2n\rceil))$ samples.}
\label{tab:numerical-checks}
\begin{tabularx}{\linewidth}{@{}>{\raggedright\arraybackslash}p{0.22\linewidth}Y@{}}
\toprule Check & Acceptance rule \\
\midrule
Basic evidence & Solver log, residual records and final-time directory present; no fatal or non-finite markers. \\
\midrule
Mesh & No failed \texttt{checkMesh} checks or severe topology/geometry markers. \\
\midrule
Final fields & Supported final-time scalar fields have parseable, finite values; \texttt{T}, \texttt{rho}, \texttt{k}, \texttt{epsilon}, \texttt{omega} and \texttt{nuTilda} $\geq-10^{-12}$; \texttt{alpha.water} and \texttt{alpha.air} in $[-10^{-3},1.001]$. \\
\midrule
Linear residuals & For initial and final residuals, largest field-wise tail p95 $\leq\max(10|x^\star|,10^{-3})$. \\
\midrule
Courant number & Tail p95 $\leq\max(3|x^\star|,1)$; historical peak $\leq\max(3|x^\star|,2)$. \\
\midrule
Continuity & Tail p95 of absolute global error $\leq\max(10|x^\star|,10^{-5})$; absolute final cumulative error $\leq\max(10|x^\star|,10^{-3})$. \\
\midrule
Steady response & In enabled FoamBench steady-state evaluations, normalized tail range $\leq\max(0.02,2s^\star)$, with reference range $s^\star$ under the same normalization. At least three samples required. \\
\bottomrule
\end{tabularx}
\end{table}
For steady-state responses, the normalized tail range is computed over the final $\max(3,\lceil0.2n\rceil)$ samples and normalized by $\max(|q_{\mathrm{last}}|,10^{-12})$. Task-level integral constraints, such as phase-volume drift or species closure, are evaluated under QoI Satisfaction when they form part of the prescribed response requirements.

\subsection{QoI Measurements and Tolerances}
\label{app:qoi-reference}
Table~\ref{tab:actual-qoi-contracts} presents representative principal QoIs and their acceptance rules. Candidate and reference quantities are extracted using the same metric definitions, units, spatial coordinates, and temporal alignment procedures. QoI Satisfaction requires all mandatory criteria associated with the task's principal QoIs to pass; auxiliary diagnostics are retained for interpretation but do not independently determine acceptance.

\begin{table}[htbp]
\centering\footnotesize
\renewcommand{\arraystretch}{1.14}
\caption{Representative principal QoIs and acceptance rules. All mandatory criteria associated with a task’s principal QoIs must pass.}
\label{tab:actual-qoi-contracts}
\begin{tabularx}{\linewidth}{@{}>{\raggedright\arraybackslash}p{0.20\linewidth}Y Y@{}}
\toprule
Task and assessed response & Measurement & Acceptance rule \\
\midrule
Basic: Cavity/1; recirculation and velocity profiles & Final velocity field and physical-coordinate centre-line profiles; maximum response speed and minimum streamwise velocity. & Speed $\in[0.1,1.0]$ m/s; reverse-flow velocity $\in[-1,-0.05]$ m/s; normalized profile $L_1\leq0.08$, $L_\infty\leq0.20$. \\
\midrule
Advanced: Double\_Square\_SA; pressure and profile response & Response profiles and the absolute difference between volume-weighted mean pressures in the first and last $\max(1,\lfloor N/10\rfloor)$ cells ordered by streamwise ($x$) coordinate, where $N$ is the total cell count. & Kinematic pressure difference $\in[0.005,10]$ m$^2$/s$^2$; profile $L_1\leq0.08$, $L_\infty\leq0.20$. \\
\midrule
Ext: yPlusVerification/6; near-wall resolution & First-cell $y^+$ from the wall-owner cell-centre distance, wall shear and kinematic viscosity. & Reference-band half-width $\max(0.1|q^\star|,0.05)$ (dimensionless). \\
\midrule
Ext: MRF\_\allowbreak stirredReactor/1; bulk-flow speed & Final cell-volume-weighted speed, $\sum_i V_i|\mathbf U_i|/\sum_iV_i$. & Relative error $\leq10\%$. \\
\midrule
Ext: aerofoilNACA0012/1; aerodynamic loads & $C_D,C_L,C_M$ from \texttt{forceCoeffs}, with task-defined surfaces, directions and normalization. & For each coefficient, half-width $\max(0.05|q^\star|,a_q)$; $a_D=5\times10^{-4}$, $a_L=10^{-3}$, $a_M=5\times10^{-4}$. \\
\midrule
Ext: nozzleCDV/1; mass-flow evolution & Declared mass-flow history, aligned to reference times without extrapolating missing support. & Normalized RMSE $\leq0.10$, maximum error $\leq0.25$, coverage $\geq90\%$ of reference samples. \\

\bottomrule
\end{tabularx}
\end{table}
Spatial profiles are compared after alignment in physical coordinates and are normalized by the larger of the mean absolute reference magnitude and a field-specific floor. Time histories are aligned to reference times by linear interpolation over the available candidate interval, without extrapolation, and require sufficient temporal coverage. Scalar quantities are evaluated using fixed reference bands with absolute floors for near-zero values.

All response intervals, normalization rules, extraction procedures, and acceptance tolerances are fixed as part of the benchmark specification before system evaluation and are applied unchanged across methods.

\subsection{Reference Reproducibility and QoI Discrimination}
\label{app:reference-validation}

We validate the reference cases and QoI evaluators using the first task (\texttt{/1}) from each of the ten ExtFoamBench families. Each task is evaluated under four conditions: the specified reference configuration, an identical repeat, a tenfold reduction in absolute linear-solver tolerances, and one deliberately perturbed physical condition (Table~\ref{tab:reference-validation}). Reference values, QoI acceptance tolerances, and measurement definitions are kept fixed across all conditions.

\begin{table}[htbp]
\centering\footnotesize
\renewcommand{\arraystretch}{1.12}
\caption{Reference-case validation across all ten families using task \texttt{/1}. B: baseline; R: identical repeat; T: tighter absolute linear-solver tolerances; N: physical negative control. P denotes acceptance and F indicates that at least one mandatory QoI falls outside its acceptance tolerance.}
\label{tab:reference-validation}
\begin{tabularx}{\linewidth}{@{}>{\raggedright\arraybackslash}p{0.25\linewidth}Y c c c c@{}}
\toprule
Family & Physical perturbation in N & B & R & T & N \\
\midrule
yPlusVerification & Kinematic viscosity multiplied by ten & P & P & P & F \\
aerofoilNACA0012 & Incidence changed from $0^\circ$ to $5^\circ$ & P & P & P & F \\
nozzleCDV & Inlet speed halved & P & P & P & F \\
risingBubble & Gravity reversed & P & P & P & F \\
supersonicCrossJet & Jet speed halved & P & P & P & F \\
MRF\_stirredReactor & Impeller speed halved & P & P & P & F \\
ERCOFTA\_pump & Prescribed inlet flow halved & P & P & P & F \\
compressible\allowbreak BubbleCollapse & Initial bubble radius doubled & P & P & P & F \\
drivaerFastback & Body-wall condition changed to slip & P & P & P & F \\
riemann2D & Prescribed quadrant velocities removed & P & P & P & F \\
\bottomrule
\end{tabularx}
\end{table}

All ten identical repeats reproduce QoIs within the original acceptance ranges. The tighter-solver-tolerance condition is also accepted for all ten families, indicating that the evaluation is stable to this numerical perturbation. In contrast, every deliberately perturbed physical condition produces at least one QoI outside its prescribed tolerance. Across the 40 comparisons, all 30 positive conditions are accepted and all ten physical negative controls are rejected.

Together, these results support the reproducibility of the selected reference cases and the ability of the frozen QoI evaluators to distinguish the tested physical perturbations under the prescribed benchmark configurations.

\subsection{Result Aggregation}
\label{app:aggregation}

For tasks requiring multiple runs, each prescribed branch is evaluated using its corresponding execution evidence and QoIs, and all mandatory branch-level criteria must pass for the task to be accepted.

A mandatory criterion is recorded as \textit{unknown} when the required evidence is unavailable or insufficient and as \textit{fail} when the available evidence demonstrates a violation. Neither outcome receives pass credit. Each task contributes one final acceptance outcome, and every assigned task remains in the denominator, including failed, timed-out, and incomplete tasks.

All three evaluation dimensions are computed from the final candidate produced by the corresponding rollout using the same frozen evaluator and aggregation rules across systems.

\subsection{Sensitivity to Acceptance Tolerances}
\label{app:threshold-sensitivity}

We reassess the frozen outputs used in Table~\ref{tab:main-results} on all 126 FoamBench and 100 ExtFoamBench tasks, without rerunning agents or solvers. The scalable error tolerances for Numerical Soundness and QoI Satisfaction are independently multiplied by
\(s_N,s_Q\in\{0.5,1,2\}\). Task-imposed hard limits, physical bounds, prescribed response intervals, reference values, normalization procedures, extraction rules, sampling windows, coverage requirements, and Execution Completion remain unchanged. The original benchmark setting corresponds to $(s_N,s_Q)=(1,1)$.

\begin{table}[htbp]
\centering\small
\caption{Accepted task counts across the complete tolerance grid. Denominators are 126 for FoamBench and 100 for ExtFoamBench; $(1,1)$ uses the original tolerances.}
\label{tab:threshold-grid}
\begin{tabular}{@{}ccrrrr@{}}
\toprule
$s_N$ & $s_Q$ & MetaOpenFOAM 2.0 & Foam-Agent & ChatCFD & EasyFoam \\
\midrule
\multicolumn{6}{l}{\textbf{FoamBench}} \\
0.5 & 0.5 & 3 & 2 & 16 & 43 \\
0.5 & 1 & 6 & 5 & 18 & 49 \\
0.5 & 2 & 8 & 7 & 20 & 57 \\
1 & 0.5 & 3 & 2 & 18 & 46 \\
1 & 1 & 6 & 5 & 21 & 52 \\
1 & 2 & 8 & 8 & 23 & 60 \\
2 & 0.5 & 3 & 2 & 18 & 48 \\
2 & 1 & 6 & 5 & 21 & 54 \\
2 & 2 & 8 & 8 & 23 & 62 \\
\midrule
\multicolumn{6}{l}{\textbf{ExtFoamBench}} \\
0.5 & 0.5 & 0 & 0 & 0 & 4 \\
0.5 & 1 & 0 & 0 & 0 & 6 \\
0.5 & 2 & 0 & 0 & 0 & 8 \\
1 & 0.5 & 0 & 0 & 0 & 4 \\
1 & 1 & 0 & 0 & 0 & 6 \\
1 & 2 & 0 & 0 & 0 & 8 \\
2 & 0.5 & 0 & 0 & 0 & 5 \\
2 & 1 & 0 & 0 & 0 & 7 \\
2 & 2 & 0 & 0 & 0 & 9 \\
\bottomrule
\end{tabular}
\end{table}

EasyFoam attains the highest TAR in all nine tolerance settings (Table~\ref{tab:threshold-grid}). The three baselines remain at zero TAR on ExtFoamBench even under the loosest joint setting, while the relative pattern on FoamBench is also stable across the grid.

To examine why relaxed tolerances recover only a limited number of tasks, Table~\ref{tab:threshold-failures} summarizes the remaining failure modes at $(2,2)$.

\begin{table}[htbp]
\centering\footnotesize
\setlength{\tabcolsep}{5pt}
\caption{Tasks that executed successfully but were not accepted at $(1,1)$. Recovered tasks become accepted at $(2,2)$; remaining tasks are still unaccepted. The final three columns give overlapping reasons among remaining tasks: missing or invalid evidence, measured failures of fixed conditions, and errors above the relaxed tolerances.}
\label{tab:threshold-failures}
\begin{tabularx}{\linewidth}{@{}l*{6}{>{\centering\arraybackslash}X}@{}}
\toprule
Method & \shortstack{Original\\failures} & Recovered & Remaining & Evidence & \shortstack{Fixed\\conditions} & \shortstack{Error\\exceedance} \\
\midrule
\multicolumn{7}{l}{\textbf{FoamBench}} \\
MetaOpenFOAM 2.0 & 30 & 2 & 28 & 11 & 7 & 19 \\
Foam-Agent & 70 & 3 & 67 & 0 & 24 & 67 \\
ChatCFD & 23 & 2 & 21 & 0 & 6 & 20 \\
EasyFoam & 71 & 10 & 61 & 2 & 12 & 60 \\
\midrule
\multicolumn{7}{l}{\textbf{ExtFoamBench}} \\
MetaOpenFOAM 2.0 & 8 & 0 & 8 & 3 & 0 & 5 \\
Foam-Agent & 27 & 0 & 27 & 15 & 1 & 19 \\
ChatCFD & 13 & 0 & 13 & 6 & 1 & 6 \\
EasyFoam & 66 & 3 & 63 & 10 & 2 & 58 \\
\bottomrule
\end{tabularx}
\end{table}

Most remaining failures still involve deviations beyond the relaxed tolerances, while missing or invalid evidence also contributes substantially on ExtFoamBench. The limited recovery across the tolerance grid indicates that the observed acceptance gap is not explained solely by the choice of the default tolerance scale.

\section{EasyFoam Implementation Details}
\label{app:easyfoam-implementation}

This section provides implementation details for the EasyFoam framework introduced in Section~\ref{sec:method}. Task-visible checks guide the execution loop, while offline evaluation supplies the outcomes used for benchmark reporting and paired skill selection.

\subsection{Canonical Records and Externalized Skill Organization}
\label{app:skill-organization}

EasyFoam maintains a shared set of canonical records to distinguish task requirements, active workflow state, execution attempts, derived outputs, and assessment evidence. These records identify the valid upstream artifacts available to each downstream operation and preserve their provenance. Table~\ref{tab:easyfoam-artifacts} summarizes their roles.

\begin{table}[htbp]
\centering
\footnotesize
\renewcommand{\arraystretch}{1.10}
\caption{Canonical records used by EasyFoam.}
\label{tab:easyfoam-artifacts}
\begin{tabularx}{\linewidth}{@{} >{\raggedright\arraybackslash}p{0.21\linewidth} >{\raggedright\arraybackslash}p{0.14\linewidth} Y @{}}
\toprule
\textbf{Record} & \textbf{Owner} & \textbf{Role} \\
\midrule
\texttt{CaseSpec} & Plan & Shared task specification containing the physical and computational setup, requested QoIs, evidence-extraction requirements, decision provenance, and unresolved assumptions. \\
\midrule
Workflow State & Workflow & Active stage, completion status, currently valid artifacts, evidence references, and recovery state. \\
\midrule
Run Record & Run & Record of one concrete execution attempt, including the executed command, environment, logs, termination status, and the case state associated with the attempt. \\
\midrule
Postprocess Results & Postprocess & Requested QoIs, sampled data, curves, tables, and other derived outputs obtained from the current simulation result. \\
\midrule
Evidence Summary & Check & Outcomes of task-visible checks together with their supporting evidence, diagnostics, and unresolved evidence requirements for the current candidate. \\
\bottomrule
\end{tabularx}
\end{table}

These records distinguish the current workflow state from historical attempts and associate artifacts with the configuration and execution attempt that produced them.

The externalized capability system introduced in Section~\ref{sec:method} comprises Hierarchical Skills, References, Schemas, and Tools. Skills follow the Root–Stage–Specialist structure. Root-level instructions route work to the execution or evolution branch. Stage-level instructions define entry, completion, handoff, and recovery conditions associated with a workflow stage. Specialist skills provide bounded CFD procedures for concrete activities such as task planning, mesh preparation, case authoring, execution, runtime diagnosis, and quantitative post-processing.

References provide read-only CFD and OpenFOAM knowledge; Schemas define the canonical records and their required fields; and Tools supply deterministic operations for state management, parsing, evidence extraction, and repeatable workflow routines. By default, only the stage-level and specialist skills, together with the supporting resources relevant to the current decision, are loaded. Shared workflow records connect these procedures while keeping the CFD guidance external to the underlying coding agent.

\subsection{CaseSpec Compilation and Evidence Planning}
\label{app:casespec-details}

The Plan stage consolidates the task request and available assets into \texttt{CaseSpec}, the central task specification used by all subsequent stages. Consistent with Section~\ref{sec:method}, \texttt{CaseSpec} defines three primary groups: the physical setup (models, properties, boundaries); the computational setup (geometry, mesh, numerical controls); and the requested QoIs coupled with their evidence-extraction requirements.

Each decision in \texttt{CaseSpec} retains an explicit provenance label:
\begin{equation}
\mathrm{prov}(d_i) \in \{ \texttt{explicit}, \texttt{derived}, \texttt{default}, \texttt{unresolved} \},
\label{eq:casespec-provenance}
\end{equation}
corresponding respectively to an explicit task requirement, a domain-supported derivation from available constraints, a permitted default for a non-decisive implementation choice, and an unresolved decision.
Derived values retain their supporting rationale, and defaults must remain consistent with the specified physical problem and requested outputs.
Unconfirmed assumptions affecting the physical problem, computational setup, or target QoIs are marked as \texttt{unresolved}. Downstream execution is blocked when such an item prevents construction of a workflow consistent with the task requirements.

Evidence requirements are planned concurrently with the computational setup. Requested QoIs dictate which quantities must be accessible for assessment, thereby constraining the monitors, sampling locations, temporal coverage, averaging intervals, and post-processing operations configured downstream. For example, a requested force coefficient requires the specification of force surfaces and normalization quantities, while a time-dependent QoI necessitates sufficient simulated temporal coverage.

\texttt{CaseSpec} contains only task requirements and information available to the solving agent. Hidden reference quantities and reference-dependent acceptance criteria are reserved for offline evaluation.

\subsection{Workflow Execution, State Handoffs, and Stage Recovery}
\label{app:workflow-recovery}

The execution loop follows the staged workflow:
\[ \textbf{Plan} \rightarrow \textbf{Write} \rightarrow \textbf{Run} \rightarrow \textbf{Postprocess}. \]
Each stage uses available upstream artifacts and produces the outputs required for subsequent operations. Table~\ref{tab:easyfoam-stages} summarizes the stage responsibilities and typical recovery conditions.

\begin{table}[htbp]
\centering
\footnotesize
\renewcommand{\arraystretch}{1.10}
\caption{Stage responsibilities and recovery boundaries in EasyFoam.}
\label{tab:easyfoam-stages}
\begin{tabularx}{\linewidth}{@{} >{\raggedright\arraybackslash}p{0.13\linewidth} >{\raggedright\arraybackslash}p{0.40\linewidth} Y @{}}
\toprule
\textbf{Stage} & \textbf{Responsibility and key outputs} & \textbf{Typical recovery condition} \\
\midrule
Plan & Establish \texttt{CaseSpec} and the execution and evidence-collection strategy. & Missing or conflicting decisive task requirements. \\
\midrule
Write & Prepare the physical models, mesh, fields, boundary conditions, numerical settings, and required runtime outputs. & Mesh, field, boundary-condition, physical-model, or numerical-configuration defects. \\
\midrule
Run & Execute the prescribed solvers and auxiliary utilities and record their outputs. & Command, restart, runtime, or solver failures requiring execution-stage correction. \\
\midrule
Postprocess & Extract requested QoIs and assemble the evidence required by task-visible checks. & Missing, incomplete, or incorrect evidence extraction. \\
\bottomrule
\end{tabularx}
\end{table}
Workflow State records stage progress and artifact validity, while the shared CaseSpec maintains the requirements that each stage must satisfy. Stage handoffs depend on realized outputs. For example, the generated mesh inventory constrains field and boundary definitions, and the available simulation times determine which QoIs can be extracted. These dependencies provide the basis for locating the stage responsible for a failed check.

During execution, EasyFoam applies task-visible checks aligned with the evaluation dimensions in Section~\ref{bench:evaluation}, using the current task requirements and available simulation evidence. A \texttt{fail} outcome records an observed violation, whereas an \texttt{unknown} outcome signals that the evidence required for a mandatory check remains unavailable. Either state triggers diagnosis.

Diagnosis pinpoints the earliest workflow stage implicated by the failure. A missing QoI routes back to Postprocess when the necessary fields exist; insufficient temporal coverage routes to Run; a configuration anomaly routes to Write; and conflicting task requirements return to Plan. The Specialist for the implicated stage then orchestrates the targeted repair.

Repair invalidates affected artifacts and their dependent downstream outputs. Unaffected artifacts may be retained when they remain consistent with the revised configuration. Historical attempts remain inspectable, but invalidated outputs cannot support assessment of the new candidate. The reopened stage and its affected downstream operations are then reassessed using valid evidence.

\subsection{Execution Provenance and Evidence Management}
\label{app:run-provenance}

EasyFoam preserves execution attempts as separate Run Records. Each record links the executed commands, working directory, environment, termination status, and logs to the associated case and mesh configuration. A repaired candidate therefore produces a new execution record while retaining the attempt that motivated the repair.

The framework distinguishes formal runs, resumed runs, smoke tests, and diagnostic runs. Formal runs and valid continuations provide evidence for assessing Execution Completion, whereas smoke tests and diagnostic runs support intermediate checks during construction and diagnosis.

\subsection{Cross-Task Skill Evolution Implementation}
\label{app:evolution-implementation}

The evolution loop revises the externalized skill instructions across task rollouts while keeping the underlying model weights fixed. Each round collects task experience, generates candidate revisions from a frozen parent library, and selects an update through paired evaluation with family-level regression penalties.

\subsubsection{Construction-task experience}

At evolution round $l$, the current skill library $\mathcal{K}_l$ is frozen and applied to a construction-task subset $\mathcal D_l^{\mathrm{con}}$. The resulting records provide the proposal agent with task requirements, workflow states, check outcomes, diagnostics, and repair traces.

Both failed and successful trajectories inform candidate updates: repeated failures identify procedures to revise, while successful trajectories provide examples to retain.

The proposal agent analyzes this aggregated experience together with the current skill instructions and proposes reusable procedural changes. For instance, recurrent field--patch mismatches motivate a mesh-inventory validation step before boundary authoring, while missing runtime observables motivate preparing output configurations during case authoring.

\subsubsection{Candidate generation and structural checks}
Each candidate skill library $\mathcal K_l^{(k)}$ is constructed as an isolated revision of the same frozen parent $\mathcal K_l$:
\begin{equation}
\mathcal K_l^{(k)} = \mathrm{Edit} \left( \mathcal K_l; \Delta_l^{(k)} \right),
\label{eq:candidate-materialization}
\end{equation}
where $\Delta_l^{(k)}$ denotes the proposed edits to the permitted files. Sequential generation allows later proposals to incorporate insights from earlier ones, while each candidate applies its edits to the same parent $\mathcal K_l$.

In these experiments, two candidates are generated per round: one prioritizes execution and numerical stability, and the other targets QoI extraction and evidence collection. Before entering simulation-based evaluation, structural checks assess readability, schema compliance, and edit bounds, filtering out malformed revisions.

\subsubsection{Paired evaluation}

The parent and structurally valid candidates are evaluated on a validation subset \(\mathcal D_l^{\mathrm{val}}\) that is disjoint from the construction subset for the current round. They use the same validation tasks, model backend, evaluation criteria, and execution budgets.

After each rollout, the offline evaluator computes the Execution Completion, Numerical Soundness, and QoI Satisfaction indicators used in the reward defined in Equation~(\ref{eq:reward}). This reward credits improvements in individual evaluation dimensions and serves as a surrogate for task acceptance; an increase in reward does not necessarily imply an increase in TAR.

Task rewards are averaged within each represented family. Candidate gains are measured relative to the parent on the same tasks, and the selection objective combines the weighted gains with a penalty for negative family-level changes. The penalty favors candidates with broader improvements but does not require every family to improve.

\subsubsection{Checkpoint replacement and subsequent rounds}

The candidate maximizing the strictly positive selection objective replaces the parent:
\begin{equation}
\mathcal K_{l+1} =
\begin{cases}
\mathcal K_l^{(k^\star)}, & k^\star = \arg\max_k J_{l,k}, \quad J_{l,k^\star}>0, \\[2mm]
\mathcal K_l, & \max_k J_{l,k}\le 0.
\end{cases}
\label{eq:app-skill-update}
\end{equation}
If no candidate yields a positive objective, the parent library is retained. The retained library is then frozen as the parent for the next round, allowing accepted skill updates to accumulate across rounds.

\section{Experimental Settings}
\label{app:reproducibility}

\subsection{Systems and Runtime Configuration}
\label{app:runtime-settings}

All systems use OpenFOAM Foundation v13 on Linux and the same offline evaluation criteria. The three external baselines request \texttt{deepseek-v4-flash}, while retaining their respective orchestration and repair mechanisms. The Coding Agent (w/o Repair \& Skills) shares EasyFoam's DeepSeek-V4-Flash backend, Codex CLI shell/file interface, and per-task rollout limit. Benchmark rollouts use fixed task prompts that provide sufficient specifications for execution, so no interactive clarification is needed.
Table~\ref{tab:runtime-settings} lists the system-specific rollout, command, and repair limits.

\begin{table}[htbp]
\centering\footnotesize
\renewcommand{\arraystretch}{1.12}
\caption{System-specific execution settings. Limits apply per task; concurrent tasks share host resources.}
\label{tab:runtime-settings}
\begin{tabularx}{\linewidth}{@{}>{\raggedright\arraybackslash}p{0.25\linewidth}Y@{}}
\toprule System & Runtime settings \\
\midrule
EasyFoam and ablation variants & 7200 s rollout limit. \\
Foam-Agent & 7200 s rollout limit; 3600 s solver timeout; up to 25 review loops. \\
ChatCFD & 1800 s rollout limit; 90 s API timeout; up to three generation attempts and three repair rounds. \\
MetaOpenFOAM 2.0 & 7200 s rollout limit; 1200 s command timeout; 120 s API timeout; up to ten repair loops within one rollout. \\
\bottomrule
\end{tabularx}
\end{table}

\paragraph{Baseline adaptation and runtime budgets.}
Baseline adaptations to OpenFOAM Foundation v13 covered version-specific guidance, tutorial resources, solver invocation, dictionary conventions, and postprocessing. The ChatCFD rollout limit applies to the agent process; 5/226 tasks reached this limit, while most terminated after completing their configured generation and repair attempts. Appendix~\ref{app:resources} reports recorded resource use.

\subsection{Ablation Conditions}
\label{app:ablation-protocol}
Ablations use the same subset of 63 FoamBench tasks (55 Basic tasks and eight Advanced tasks) for all four variants in Table~\ref{tab:component-ablation}. All four variants are evaluated using the same task-specific criteria and aggregation rules. Table~\ref{tab:ablation-policy-details} specifies these conditions.

\begin{table}[htbp]
\centering\footnotesize
\caption{Component settings for the four variants in the ablation study.}
\label{tab:ablation-policy-details}
\begin{tabularx}{\linewidth}{@{}>{\raggedright\arraybackslash}p{0.27\linewidth}Y@{}}
\toprule Method variant & Configuration \\
\midrule
EasyFoam & Retain domain skills, workflow guidance and task-visible checks. Use check findings to diagnose defects, reopen the implicated stage, repair the case and regenerate affected evidence. \\
\midrule
w/o Stage-Scoped Repair & Retain domain skills, workflow guidance, task-visible checks and retry. Replace stage-scoped diagnosis, dependent-evidence invalidation and targeted repair with generic whole-case retry. \\
\midrule
w/o Repair & Retain domain skills, workflow guidance and task-visible checks, but report check findings without using them to correct and re-execute the case. \\
\midrule
w/o Repair \& Skills & Further remove domain skills and their associated workflow guidance from the w/o Repair variant. \\
\bottomrule
\end{tabularx}
\end{table}
Recovery policies are specified through instructions; the editing and execution tools remain available. The w/o Stage-Scoped Repair variant replaces stage-scoped diagnosis, dependent-evidence invalidation and targeted repair as a combined recovery policy. Within each track, this variant and full EasyFoam share the task prompts, skill snapshot, model, OpenFOAM version, evaluator and per-task rollout limit.

\subsection{Skill Evolution and Test Settings}
\label{app:evolution}
\label{app:experience}
\label{app:candidate}
\label{app:candidate-validation}
\label{app:evolution-governance}
\label{app:evolution-experiment}
Table~\ref{tab:app-evolution-settings} specifies the skill-evolution experiment. Its development pool contains 54 Basic, eight Advanced and 50 ExtFoamBench tasks.
Each round draws two disjoint ten-task subsets from the development pool: one supplies experience for candidate generation, and the other is reserved for paired evaluation. Both subsets follow the 4/1/5 allocation across Basic/Advanced/ExtFoamBench and are selected by deterministic family/variant rotation. Two proposals start from the same parent, emphasizing execution/numerics and QoI/evidence respectively. Edits are confined to designated skill files, and the parent evaluation is shared across candidate comparisons.

\begin{table}[htbp]
\centering\footnotesize
\renewcommand{\arraystretch}{1.12}
\caption{Skill-evolution configuration.}
\label{tab:app-evolution-settings}
\begin{tabularx}{\linewidth}{@{}>{\raggedright\arraybackslash}p{0.30\linewidth}Y@{}}
\toprule Setting & Value \\
\midrule
Model and reasoning & DeepSeek-V4-Flash, \texttt{max}, for proposals and task execution. \\
\midrule
Run length & Three rounds; two proposals per round; stop at the configured round limit. \\
\midrule
Proposal budget & 3600 s per candidate; at most two permitted skill files or local edit blocks. \\
\midrule
Task-execution budget & 7200 s per rollout; mesh guard of four million cells. \\
\midrule
Reward weights & $(\alpha_E,\alpha_N,\alpha_Q)=(0.2,0.3,0.5)$. \\
\midrule
Selection settings & Ten paired-validation tasks per round; equal weights across families represented in that round's validation set; regression-penalty coefficient $\lambda=0.5$. \\
\bottomrule
\end{tabularx}
\end{table}
Initial denotes the skills before evolution; Evolved denotes the final skills frozen after evolution. Both are evaluated on the same 57 held-out test tasks: 28 Basic, four Advanced and 25 ExtFoamBench. These tasks are excluded from candidate generation and selection. Families are represented in both development and test sets, so this experiment measures transfer to held-out tasks within the covered families. Both conditions use DeepSeek-V4-Flash with \texttt{max} reasoning, the same task prompts and evaluation criteria, and a 7200 s per-task rollout limit.

\section{Case Studies and Resource Use}
\label{app:supplementary-results}

\subsection{Construction and Repair Records for the Ablation Study}
\label{app:repair-evidence}

\paragraph{Requirement implementation: \texttt{squareBend/4}.}
We examine two independent rollouts of \texttt{squareBend/4}, one with \textbf{w/o Repair} and one with \textbf{EasyFoam}. Both retain domain skills and workflow guidance. Table~\ref{tab:squarebend-requirement-mapping} records selected task requirements and their implementation in both generated cases. Both cases use \texttt{steadyState} time discretization, so the runtime coordinate denotes steady-state iterations.

\begin{table}[t]
\caption{Selected requirement-to-configuration mappings shared by the two independent \texttt{squareBend/4} rollouts. Paths are relative to the generated case directory.}
\label{tab:squarebend-requirement-mapping}
\centering
\footnotesize
\setlength{\tabcolsep}{3pt}
\renewcommand{\arraystretch}{1.12}
\begin{tabularx}{\linewidth}{@{}
>{\raggedright\arraybackslash}p{0.25\linewidth}
>{\raggedright\arraybackslash}p{0.34\linewidth}
>{\raggedright\arraybackslash}X@{}}
\toprule
User requirement & File / entry & Generated setting \\
\midrule
Inlet mass flow: $0.5\,\mathrm{kg\,s^{-1}}$
& \texttt{0/U}: inlet / \texttt{massFlowRate}
& \texttt{constant 0.5} \\

Inlet profile: \texttt{turbulentBL}
& \texttt{0/U}: inlet / \texttt{profile}
& \texttt{turbulentBL} \\

Dynamic viscosity: $2.5\!\times\!10^{-5}\,\mathrm{Pa\,s}$
& \texttt{constant/}\newline\texttt{physicalProperties}: \texttt{mixture/transport/mu}
& \texttt{2.5e-05} \\

Prandtl number: $0.71$
& \texttt{constant/}\newline\texttt{physicalProperties}: \texttt{mixture/transport/Pr}
& \texttt{0.71} \\

Start, endpoint and step: $0$, $500$, $1$
& \texttt{system/controlDict}: \texttt{startTime}, \texttt{endTime}, \texttt{deltaT}
& \texttt{0}; \texttt{500}; \texttt{1} \\

Output interval: $100$
& \texttt{system/controlDict}: \texttt{writeControl}, \texttt{writeInterval}
& \texttt{timeStep}; \texttt{100} \\
\bottomrule
\end{tabularx}
\end{table}

\paragraph{Early termination and repair: \texttt{squareBend/4}.}
Both rollouts initially terminated at iteration 145 with solver exit code zero. The solver log reported:
\begin{quote}
\small\ttfamily
PIMPLE solution converged in 145 iterations
\end{quote}
The diagnosis traced this early termination to the tutorial's \texttt{residualControl} block in \texttt{system/fvSolution}, with thresholds of $10^{-3}$ for \texttt{p}, $10^{-4}$ for \texttt{U}, and $10^{-3}$ for \texttt{h}, \texttt{k} and \texttt{epsilon}. This convergence-based stop took effect before \texttt{endTime=500}, leaving the requested output schedule incomplete.

In \textbf{w/o Repair}, the agent identified the stopping criterion but left the case unchanged; the task was not accepted. In the independent \textbf{EasyFoam} rollout, the agent reopened case authoring, removed the \texttt{residualControl} block, revalidated the configuration, and reran the solver. This repair preserved the mesh and physical boundary conditions. The revised run reached iteration 500, produced outputs at 100-iteration intervals through 500, and passed Execution Completion, Numerical Soundness and QoI Satisfaction, yielding task acceptance.

\paragraph{Run-stage repair versus whole-case retry: \texttt{shallowWaterWithSquareBump/1}.}
\label{app:stage-scoped-repair-evidence}
We compare independent full EasyFoam and w/o Stage-Scoped Repair rollouts of the same task. In each initial formal attempt, \texttt{blockMesh}, \texttt{setFields} and \texttt{shallowWaterFoam} returned zero, and the solver log contained the requested terminal time followed by \texttt{End}. The time line had a leading space, \verb| Time = 100|, which the terminal-time detector did not recognize. Both formal records were therefore \texttt{blocked}, with the reason: ``Solver returned zero but no verifiable `Time =' terminal value was found.''

The w/o Stage-Scoped Repair rollout identified the mismatch as a Run-tooling detection gap. It restored the initial fields and repeated the complete command chain without changing the detection path. The retry again reached the requested endpoint but produced the same blocked record. The agent's final report acknowledged this limitation and asserted completion from direct log inspection; the formal record remained unresolved.

Full EasyFoam created a workspace wrapper that invoked the original solver and removed leading whitespace from its output. It recorded the reason for overriding the Run-stage command and re-executed the formal command chain. The resulting record was \texttt{succeeded}, with \texttt{completed\_time\_or\_iteration=100}, and supported the subsequent runtime-evidence extraction. The blocked attempt remained available alongside the successful record.

\begin{table}[htbp]
\centering\footnotesize
\caption{Observed recovery actions for \texttt{shallowWaterWithSquareBump/1}. The comparison concerns the execution-evidence interface; both solvers reached the requested endpoint.}
\label{tab:stage-scoped-repair-trace}
\begin{tabularx}{\linewidth}{@{}>{\raggedright\arraybackslash}p{0.20\linewidth}YY@{}}
\toprule
Evidence & w/o Stage-Scoped Repair & Full EasyFoam \\
\midrule
Diagnosed defect & Indented solver time lines were not recognized by the terminal detector. & Same log-format mismatch. \\
Recovery action & Restore initial state and repeat the complete execution chain. & Normalize solver-log whitespace through a documented Run-stage wrapper, then re-execute. \\
Recorded inputs & Case-input and mesh hashes unchanged across the retry. & Case-input and mesh hashes unchanged across the repair. \\
Formal record & Remains \texttt{blocked} despite reaching the endpoint. & Becomes \texttt{succeeded}, with the endpoint verified. \\
\bottomrule
\end{tabularx}
\end{table}

The unchanged hashes confirm that recorded case inputs and mesh content were preserved within each rollout. Full EasyFoam corrected the detection mismatch and re-executed preprocessing and the solver, restoring verifiable completion. In w/o Stage-Scoped Repair, the blocked execution record prevented downstream Numerical and QoI checks, leaving those outcomes unassessed.

\subsection{Skill Updates}
\label{app:evolution-evidence}
The following development and test-task records illustrate how skill updates specify diagnostic actions, output preparation, and measurement checks.

\paragraph{Ordered diagnosis: nozzleCDV/4.}
Three back-pressure branches terminated with negative temperatures; one log reported \texttt{Negative initial temperature T0: -13401.875}. The authoring update added a coupled check of the energy variable, equation of state, pressure, density and temperature/enthalpy. The debug-reviewer update specified: ``identify the first nonphysical field, cell/time, and governing energy variable before changing numerics.'' It then required checking thermodynamic and total/static boundary consistency against Courant history, followed by testing one modification at a time in a diagnostic copy: startup timestep/Courant limit, bounded energy flux, then the implicated boundary, property or mesh condition. This record connects the nozzle failure to a reusable diagnostic sequence.

\paragraph{Output preparation: shallowWaterWithSquareBump/5.}
The authoring update replaced conditional preparation of function objects with the instruction to ``Materialize every function object that the released CaseSpec output/QoI contract depends on before the case is hashed''. It required binding each producer to the actual fields, patches and regions, and setting a cadence covering the requested measurement window. The runner also checked output readiness before formal execution. In a subsequent shallow-water rollout, the agent ran a smoke test on a separate copy and read the generated \texttt{.dat} files, including water-depth and integrated-depth histories. The trajectory recorded: ``Smoke chain works end-to-end with all 8 function objects. Launching the full 100 s formal attempt.''

\paragraph{Sampling verification: BernardCells/10.}
The postprocessing update required deriving the measurement interval from the quantity's history, checking sampling frequency, and independently verifying extraction. Insufficiently sampled measurements were to remain unresolved and trigger a bounded continuation or re-extraction. In the evolved rollout, the agent investigated an apparent heat-flux imbalance and found that heat flux updated every $50\,\mathrm{s}$ while mean temperature updated every second. It resumed from a saved state in a diagnostic run and generated one-second heat-flux records over $500$--$1000\,\mathrm{s}$.

We independently recomputed the mean of $(Q_{\mathrm{floor}}+Q_{\mathrm{ceiling}})/Q_{\mathrm{floor}}$, retaining the recorded signs, using either all samples or every fiftieth sample of this diagnostic trajectory. This statistic measures the relative imbalance between the two wall heat rates; it does not include changes in stored energy. Both calculations include the interval endpoints (Table~\ref{tab:skill-sampling-evidence}). Subsampling alone increased the magnitude of the apparent imbalance, showing that sparse records can exaggerate the discrepancy used for diagnosis. This comparison isolates the effect of sampling frequency on the reported imbalance.

\begin{table}[htbp]
\centering\small
\caption{Sampling comparison on the same \texttt{BernardCells/10} diagnostic trajectory and measurement interval.}
\label{tab:skill-sampling-evidence}
\begin{tabular}{@{}lcc@{}}
\toprule
Sampling interval & Samples & Mean relative imbalance \\
\midrule
$50\,\mathrm{s}$ & 11 & $-1.634\%$ \\
$1\,\mathrm{s}$ & 501 & $-0.638\%$ \\
\bottomrule
\end{tabular}
\end{table}

\subsection{Resource Accounting}
\label{app:resources}

\begin{table}[htbp]
\centering\small
\caption{Resources per assigned task for the indicated evaluation sets. Token coverage reports tasks with recorded usage. Daggered ranges reflect unknown input-cache splits.}
\label{tab:resource-cost}
\begin{tabular}{@{}llrrr@{}}
\toprule
Track & System & Token coverage & \shortstack{Estimated USD/\\assigned task} & \shortstack{Recorded min/\\assigned task} \\
\midrule
Basic & MetaOpenFOAM 2.0 & 110/110 & 0.019--0.051$^{\dagger}$ & 4.0 \\
& Foam-Agent & 88/110 & 0.340 & 57.8 \\
& ChatCFD & 110/110 & 0.023 & 1.6 \\
& EasyFoam & 109/110 & 0.182 & 13.8 \\
\midrule
Advanced & MetaOpenFOAM 2.0 & 16/16 & 0.021--0.056$^{\dagger}$ & 3.6 \\
& Foam-Agent & 13/16 & 0.284 & 58.0 \\
& ChatCFD & 16/16 & 0.018 & 1.0 \\
& EasyFoam & 16/16 & 0.270 & 21.7 \\
\midrule
ExtFoamBench & MetaOpenFOAM 2.0 & 100/100 & 0.025--0.064$^{\dagger}$ & 6.0 \\
& Foam-Agent & 83/100 & 0.582 & 83.4 \\
& ChatCFD & 100/100 & 0.018 & 2.3 \\
& EasyFoam & 90/100 & 0.338 & 45.4 \\
\midrule
\multicolumn{5}{l}{\textbf{Coding Agent: ablation subset}} \\

FoamBench (63) & Coding Agent & 61/63 & 0.228 & 27.5 \\
\bottomrule
\end{tabular}

\end{table}

Table~\ref{tab:resource-cost} reports resource use for the main-comparison campaigns and the Coding Agent ablation subset. Coding Agent uses the same 63 FoamBench tasks as Table~\ref{tab:component-ablation}; the remaining rows report the full indicated tracks. Costs and rollout times are averaged over all assigned tasks, including failures, timeouts, and recorded retries. Costs cover recorded token usage; missing usage is not imputed.

Costs use fixed rates of USD 0.006, 0.30 and 1.20 per million cached-input, uncached-input and output tokens, respectively. They exclude infrastructure, offline evaluation, skill development and offline embedding costs.

MetaOpenFOAM 2.0 resources correspond to the main-comparison campaigns, with token totals and rollout times recorded for all 226 tasks. The input-cache split is unavailable for part of the recorded usage. Daggered ranges therefore price recorded input tokens as fully cached or fully uncached, with output-token costs fixed.

\end{document}